\documentstyle[twoside,fleqn,espcrc2]{article}

\include{psfig}

\newcommand{\AmS}{{\protect\the\textfont2
  A\kern-.1667em\lower.5ex\hbox{M}\kern-.125emS}}

\begin{document}


\begin{titlepage}
\thispagestyle{empty}

\begin{minipage}{13.75cm}

\vspace{-1.75cm}

\begin{flushright}
   HLRZ 92-70  \\
   October 23, 1992
\end{flushright}

\begin{center}
   \begin{LARGE}
      {\bf{Renormalisation of lattice currents and the \\[0.25em]
           calculation of decay constants for          \\[0.5em]
           dynamical staggered fermions}}%
    \end{LARGE} \\[0.50cm]
    --- The $MT_c$ Collaboration ---  \\[0.50cm]
    R. Altmeyer$^{a}$, K. D. Born$^b$, M. G\"ockeler$^{b,c}$,      \\
    R. Horsley$^{c}$, E. Laermann$^{d}$ and G. Schierholz$^{a,c}$  \\[1.5em]
    {\small $^a$Deutsches Elektronen-Synchrotron DESY,}            \\[-0.25em]
    {\small Notkestra{\ss}e 85, W-2000 Hamburg 52, Germany}        \\
    {\small $^b$Institut f\"ur Theoretische Physik, RWTH Aachen,}  \\[-0.25em]
    {\small Sommerfeldstra{\ss}e, W-5100 Aachen, Germany}          \\
    {\small $^c$H\"ochstleistungsrechenzentrum HLRZ,}              \\[-0.25em]
    {\small c/o Forschungszentrum J\"ulich,
             W-5170 J\"ulich, Germany}                             \\
    {\small $^d$Fakult\"at f\"ur Physik, Universit\"at Bielefeld,} \\[-0.25em]
    {\small Postfach 8640, W-4800 Bielefeld 1, Germany}
\end{center}

\vspace{2cm}

\centerline{\bf Abstract}
\begin{quote}
A numerical calculation of the lattice staggered renormalisation constants
at $\beta = 5.35$, $m = 0.01$ is presented. It is seen that there are
considerable non-perturbative effects present. As an application
the vector decay constant $f_\rho$ is estimated.
\end{quote}

\end{minipage}

\end{titlepage}

\message{QCD preprint, October 23, 1992}

\newpage


\title{Renormalisation of lattice currents and the calculation
       of decay constants for dynamical staggered fermions%
           \thanks{The $MT_c$ Collaboration.}%
           \thanks{Talk presented by R. Horsley at the International
                   Conference on Lattice Field Theory, Amsterdam, 1992.}}
\author{R. Altmeyer%
           \address{Deutsches Elektronen-Synchrotron DESY, \\
                    Notkestra{\ss}e 85, W-2000 Hamburg 52, Germany}
        K.~D. Born%
           \address{Institut f{\"u}r Theoretische Physik, RWTH Aachen, \\
                    Sommerfeldstra{\ss}e, W-5100 Aachen, Germany}
        M. G{\"o}ckeler$^{{\rm b},}$%
           \address{H{\"o}chstleistungsrechenzentrum HLRZ,  \\
                    c/o Forschungszentrum J{\"u}lich, W-5170 J{\"u}lich,
                                                             Germany}
        R. Horsley$^{\rm c}$
        E. Laermann%
           \address{Fakult{\"a}t f{\"u}r Physik, Universit{\"a}t Bielefeld, \\
                    Postfach 8640, W-4800 Bielefeld, Germany}
        and
        G. Schierholz$^{{\rm a,c}}$}


\begin{abstract}
A numerical calculation of the lattice staggered renormalisation constants
at $\beta = 5.35$, $m = 0.01$ is presented. It is seen that there are
considerable non-perturbative effects present. As an application
the vector decay constant $f_\rho$ is estimated.
\end{abstract}

\maketitle

\setcounter{footnote}{0}

\section{Introduction}

QCD lattice calculations are now entering a phase where attempts
are being made to quantitatively understand the theory. One essential
component is the renormalisation of lattice operators, so as to
be able to compare various matrix elements with their experimental
counterparts. Although these renormalisation constants can be
computed perturbatively, it is not so clear whether with present
day couplings we are in such a perturbative regime. In this note we shall
report preliminary results%
\footnote{A fuller account will appear in \cite{altmeyer92b}.}
for a non-perturbative calculation for dynamical staggered fermions from the
$MT_c$ project: $85$ configurations were generated at $\beta = 5.35$,
$m =0.01$ on a $16^3 \times 24$ lattice \cite{altmeyer92a},
using the HMC algorithm. Staggered fermions have a natural advantage
to Wilson fermions as they have a residual $U(1) \otimes U(1)$
chiral symmetry on the lattice, and thus certain currents suffer
no renormalisation. We shall show that for the other currents,
results can be obtained that are very different from what we would
expect from perturbation  theory. As an immediate application
we shall apply the results to the calculation of the $\rho$ decay constant.

\section{Staggered Currents}

The continuum counterparts of staggered currents are given by
\begin{equation}
   J^F \sim \bar{q} \gamma_J \otimes \xi_F q
\label{currents1}
\end{equation}
where $\gamma_J$ is a Dirac matrix representing the spin structure,
which can be either $\gamma_J = 1$, $\gamma_\mu$, $-i\gamma_\mu\gamma_\nu$,
$i\gamma_\mu\gamma_5$ or $\gamma_5$ for scalar ($S$), vector ($V$),
tensor ($T$), axial ($A$) and pseudoscalar ($P$) currents respectively%
\footnote{For lack of a better convention we shall call all such operators
``currents''. We shall not consider the tensor currents here.}.
$\xi_F$ is a Dirac matrix in flavour space with $F =S$, $V$, $T$, $A$ or $P$.
(Remember that staggered fermions describe $4$ degenerate quarks
transforming under the flavour group $SU_F(4)$.) On the lattice
these currents are built up from one component Grassmann
fields: $\chi$, $\bar{\chi}$ (upon suppressing the additional colour index).
Spin and flavour degrees of freedom are distributed over the lattice sites.
A simple prescription \cite{golterman84a} is to associate
\begin{eqnarray}
   \gamma_\mu &\leftrightarrow& \eta_\mu(x) \Delta_\mu(x,y)
                                                      \nonumber \\
   \xi_\alpha &\leftrightarrow& \zeta_\alpha(x) \Delta_\alpha(x,y)
                                                      \nonumber \\
   \gamma_5 \otimes \xi_5
              &\leftrightarrow& \epsilon(x)
\label{currents2}
\end{eqnarray}
where
$\Delta_\lambda(x,y) = (\delta_{x+\hat{\lambda},y}
+ \delta_{x-\hat{\lambda},y}) / 2$
and $\eta$ and $\zeta$ are the standard staggered phase factors.
This gives the lattice current $J_0^F \sim \bar{\chi} {\cal K} \chi$
where the kernel ${\cal K}$, depending on the current considered, is
composed of $0$ to $5$ link operators. To achieve gauge invariance
we average over all the minimal paths from $\bar{\chi}$ to $\chi$.
Note that these operators can be larger than the hypercube: the motivation
is that they lead to ``simple'' Ward Identities \cite{smit88a}.
(Our numerical results will, however, only consider such operators that can
be fitted onto the hypercube.)

\section{Renormalisation}

We set
\begin{equation}
   J^F = Z_J \kappa_J^F J_0^F
\label{renormalisation1}
\end{equation}
where $J^F$ is the renormalised continuum current and $J^F_0$ is the
bare lattice current. $Z_J$ is the usual continuum renormalisation factor,
which may be divergent as $a \to 0$. The additional finite
renormalisation constant, necessary for covariance of the renormalised
currents under flavour transformations is denoted by $\kappa_J^F$.
It is now our task to calculate these numbers, which in the continuum,
by definition, are unity. We would expect that in the scaling region
flavour symmetry is restored (see e.g. \cite{altmeyer92a}) up to
$O(a^2)$ corrections \cite{sharpe92a}, while the corrections to
$\kappa$ should be $O(g^2) \sim O(1/\log a)$, and so $\kappa \to 1$
rather slowly in the (asymptotic) scaling region. (The perturbative
corrections for $\kappa_S^F$ have been calculated in \cite{golterman84b}
while a more complete set is given in \cite{daniel88a}.) Although we are
now seeing clear signs of flavour symmetry restoration, it is not so clear
if we can also expect perturbative results to hold for the renormalisation
constants.

\section{The Method}

We follow the method proposed by Smit and Vink \cite{smit88a}, which notes
that continuum matrix elements should be independent of flavour
transformations for all members of an $SU_F(4)$ multiplet
(we have a $1$ dimensional singlet and a $15$ dimensional
adjoint representation present). Thus from eq.~\ref{renormalisation1}
for two members $F$, $F^\prime$ of the same multiplet we can form ratios
to give
\begin{equation}
   \kappa_J^F = \kappa_J^{F^\prime}
                   \left| {\langle 0 | J_0^F | state \rangle} \over {
                          {\langle 0 | J_0^{F^\prime} | state \rangle} }
                   \right|
\label{method1}
\end{equation}
So given one $\kappa$ we can form a chain to calculate the others.
By convention we set $\kappa_S^S$, $\kappa_P^P$ equal to one, and as
noted previously, the currents $V^S$, $A^P$ are protected by the
chiral symmetry on the lattice, so that $\kappa_V^S = 1 = \kappa_A^P$.
We now have our `start-up' $\kappa$ values. The lattice matrix elements
$\langle 0 | J_0^F | state \rangle$ can easily be related to
the amplitude%
\footnote{Roughly speaking the amplitude is the square of the
matrix element, \cite{altmeyer92a}.}
of certain meson correlation functions, as many of the meson
operators in the Golterman Tables \cite{golterman86a}
are equivalent to the currents defined here. Thus we can now,
in principle at least, calculate the finite lattice renormalisation constants.

\section{Results}

\begin{figure}[tb]
\vspace{-1.25cm}
\centerline{\psfig{file=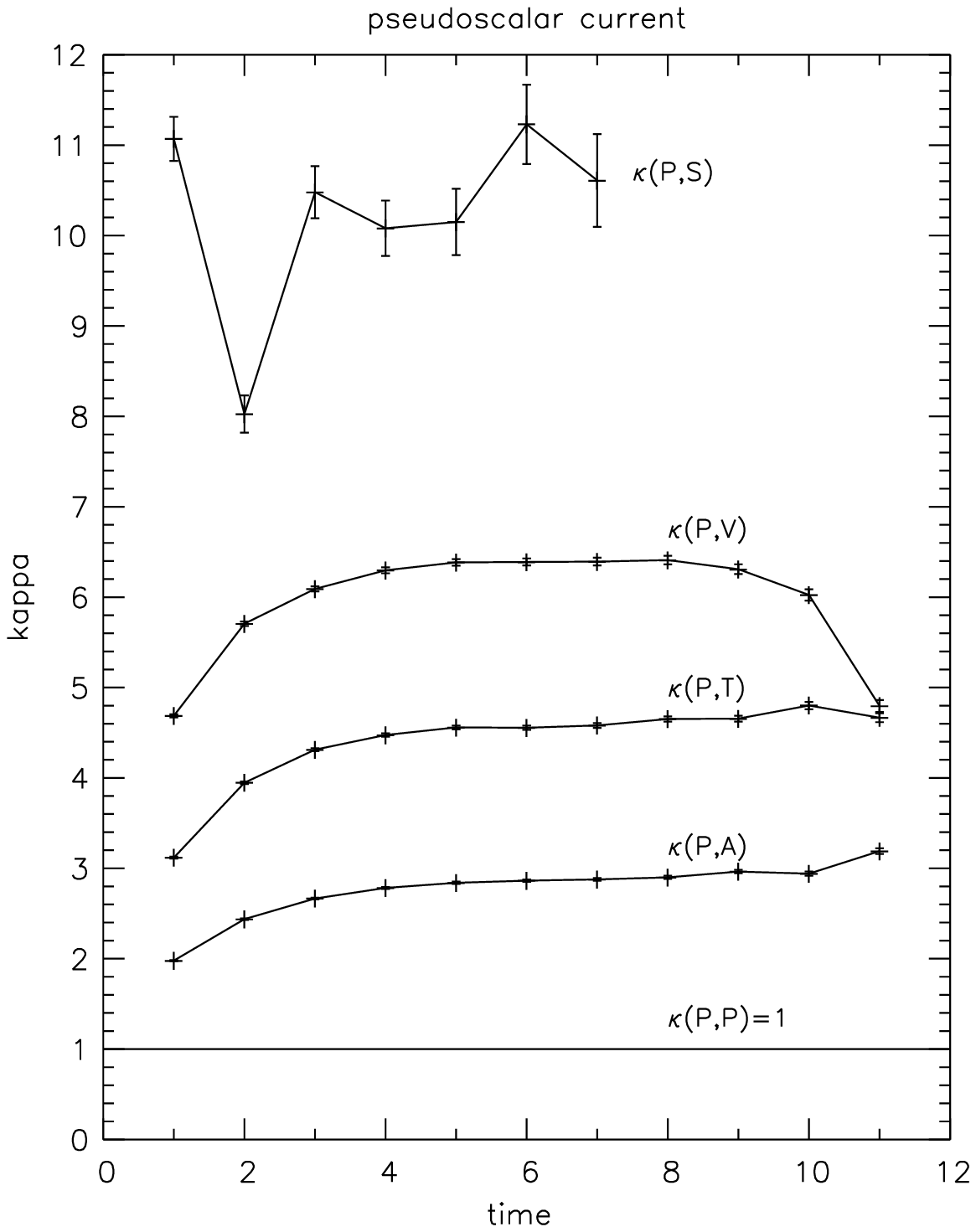,height=10cm}}
\vspace{-1.25cm}
\caption{The effective $\kappa_P^F$ for $F = P$, $A$, $T$, $V$ and $S$.}
\label{figp}
\end{figure}

\begin{figure}[tb]
\vspace{-1.25cm}
\centerline{\psfig{file=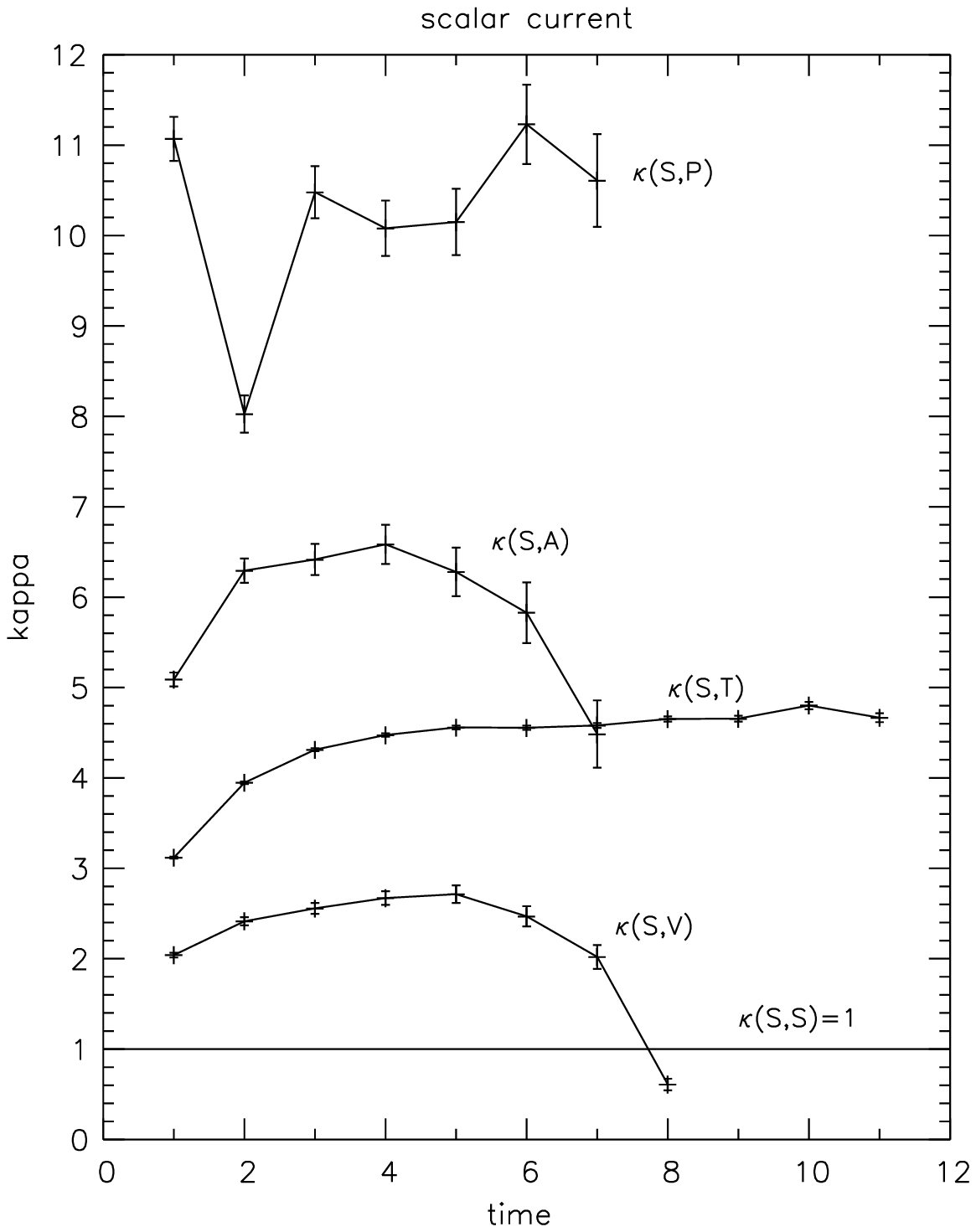,height=10cm}}
\vspace{-1.25cm}
\caption{The effective $\kappa_S^F$ for $F = P$, $A$, $T$, $V$ and $S$.}
\label{figs}
\end{figure}

We shall now briefly discuss the pseudoscalar and scalar renormalisations
and refer the reader to \cite{altmeyer92b} for the vector and axial
cases. In figs.~\ref{figp} and \ref{figs} we plot the effective $\kappa_P^F$
and $\kappa_S^F$ respectively. By effective we mean that from the fit
to the time interval $[t,24-t]$ we extract the appropriate amplitude,
which leads via eq.~\ref{method1} to the effective $\kappa(t)$.
We then look for the plateau. As discussed previously we use in
fig.~\ref{figp} $\kappa_P^P = 1$ as normalisation and as $P^P$,
$P^A$, $P^T$ and $P^V$ transform according to the $\pi$ adjoint multiplet,
we can calculate $\kappa_P^A$, $\kappa_P^T$ and $\kappa_P^V$.
Under a chiral transformation we can rotate $P^T$, $P^T_0$ to $S^T$,
$S_0^T$ respectively, and hence we have $\kappa_S^T \equiv \kappa_P^T$.
Performing the jump to fig.~\ref{figs}, then $\kappa_S^T$ can be used
to calculate $\kappa_S^V$, $\kappa_S^A$ and $\kappa_S^P$ as the
appropriate currents all transform as an $a_0$ multiplet.
Finally we can again chirally rotate $S^P$ to $P^S$ to obtain
$\kappa_P^S \equiv \kappa_S^P$. Unfortunately this last step is not
possible for us as we have not calculated the necessary correlation
function for $\langle 0 | S_0^P | a_0 \rangle$. We thus have to take
a short-cut, and simply say that as we are only calculating the
fermion connected correlations, then $P^S$ which
should transform as a singlet ($\eta^\prime$) also transforms as
a member of the $\pi$ multiplet. This then gives $\kappa_P^S$ and
hence $\kappa_S^P$. While this is not strictly true, we note that for
these calculations the $\eta^\prime$ mass is degenerate with the $\pi$ mass
\cite{altmeyer92a} and also that the final result fits into the
proceeding pattern: the larger the current operator the larger the
corresponding $\kappa$ is.

Turning now to the numerical values%
\footnote{Tables of the results will be given in \cite{altmeyer92b}.}
of the renormalisation constants, the most important point to notice is
their magnitude, the largest being $\kappa_P^S$ (or $\kappa_S^P$)
which is of $O(10)$ and is far away from from what we would expect from
perturbation theory ($\kappa < 2$). There is one additional piece of
numerical evidence for $\kappa_P^S$, which supports our claims. In
\cite{vink88a} Vink estimated this renormalisation constant in the
quenched case by generating configurations with a given topological
charge and then computing the pseudoscalar current.
(Due to the axial anomaly these quantities can be related to each other.)
He found a result for $\beta = 5.7$ of $\kappa_P^S \sim 9.2$, which
drops at $\beta = 6.3$ to $4.1$. The first number is in agreement with
our result. His results also indicate that with relatively small increases
in $\beta$, the large numerical values found here might decrease reasonably
rapidly.

For the axial and vector currents a similar procedure leads to estimates
of $\kappa_A^F$, $\kappa_V^F$. The results will be tabulated in
\cite{altmeyer92b}. We just note here that the numerical values are
a factor two or more smaller than for the pseudoscalar and scalar
currents, for example $\kappa_V^T \sim O(2)$.

\section{The vector decay constant}

\begin{figure}[tb]
\vspace{-1.25cm}
\centerline{\psfig{file=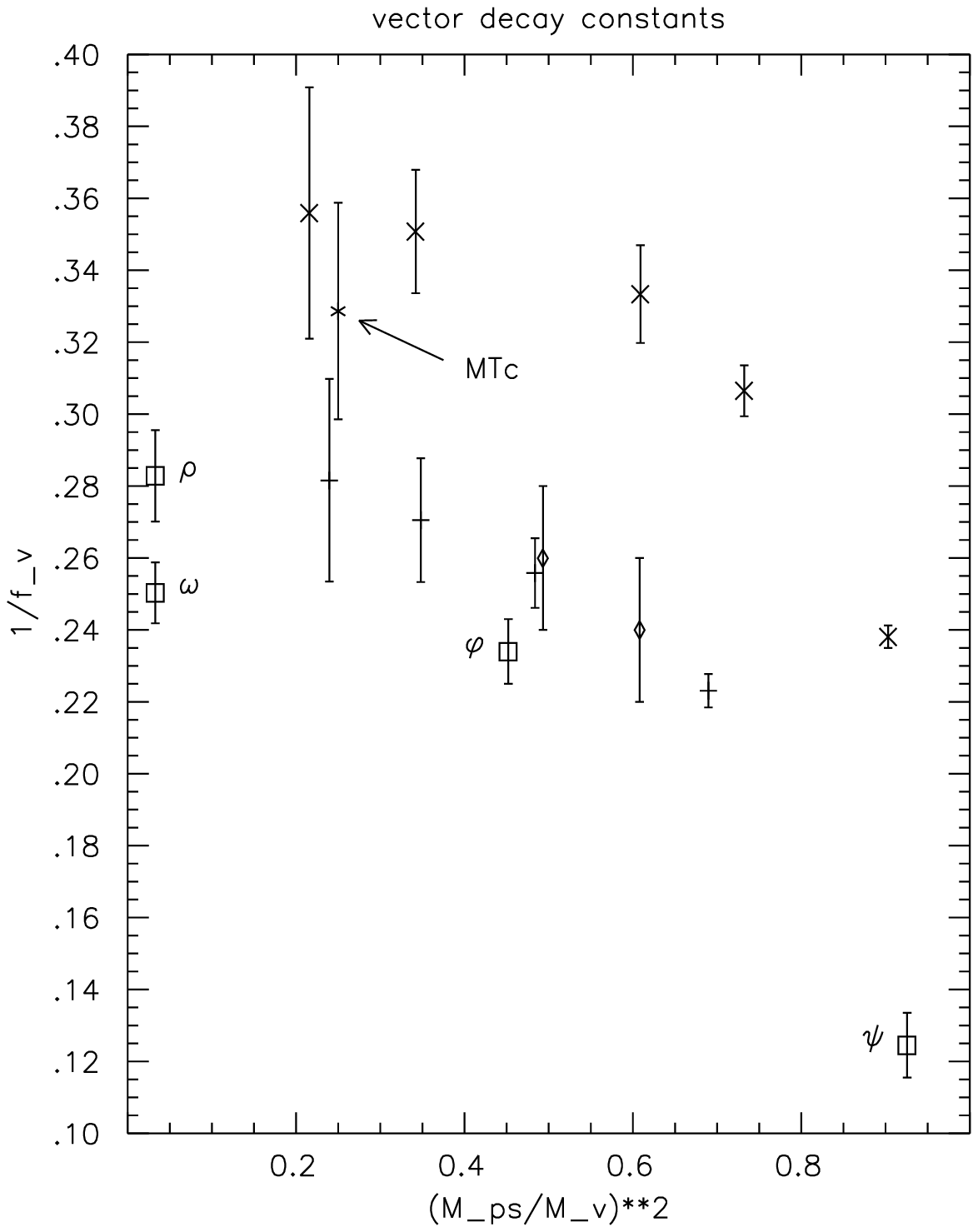,height=10cm}}
\vspace{-1.25cm}
\caption{The vector decay constant $1/f_v \equiv 2^{1/2}/f_\rho$
         is plotted against $(m_{ps}/m_v)^2$
         where $ps = \mbox{pseudoscalar}$, $v = \mbox{vector}$.}
\label{figfv}
\end{figure}

A knowledge of the dimensionless $\rho$ decay constant $f_\rho$ is
necessary for computing the amplitude of the electromagnetic
$\rho \to e^+ e^-$ decay. In the continuum $f_\rho$ is defined by
\begin{equation}
   {{\sqrt{2}m_\rho^2}\over f_\rho} \epsilon_i
          = | \langle 0 | V_i^F | \rho \rangle |
\label{decay1}
\end{equation}
where $\epsilon_i$ is the polarisation vector. On the lattice for staggered
fermions modifying the prescription given in \cite{kilcup87a} gives
\begin{equation}
   {{\sqrt{2}M_\rho^2}\over f_\rho} \epsilon_i
          = {1\over \sqrt{n_f}} \kappa_V^F | \langle 0 | V_i^F |
                                           \rho \rangle | \sqrt{N_\rho}
\label{decay2}
\end{equation}
($F \not= S$) where $M_\rho = 2\sinh (m_\rho/2) \sim m_\rho$ and $n_f = 4$.
Due to our lattice normalisation $\langle \rho | \rho \rangle =1$,
we must include an extra normalisation factor
$N_\rho = 2\sinh m_\rho \sim 2m_\rho$. Thus in eq.~\ref{decay2}
choosing an appropriate $F$ (e.g. $F = T$) we know all the terms in
the equation. This leads to $1/f_\rho = 0.232(21)$. In Fig.~\ref{figfv}
we plot the numerical and experimental results for
$1/f_v \equiv \sqrt{2}/f_\rho$.
Our result is indicated with a small cross. Also shown are results
using quenched Wilson fermions: crosses from \cite{guesken89a},
plusses from (corrected) \cite{cabasino91a} and diamonds from
\cite{daniel92a}. The experimental results are denoted by boxes.
We see that our result is a little higher than it should be, but
nevertheless we find it quite encouraging. Finally we note that
an equivalent procedure is used for the $\pi$ decay constant,
$f_\pi$. A knowledge of $\kappa_A^F$ is thus required. However as
previously discussed for staggered fermions we have $\kappa_A^P = 1$
which is invariably used.

\section{Acknowledgments}

This work was supported in part by the Deutsche Forschungsgemeinschaft.
The numerical computations were performed on the Cray Y-MP in J{\"u}lich
with time granted by the Scientific Council of the HLRZ. We thank both
institutions for their support. We would also like to thank
S.~R. Sharpe and particularly J.~C. Vink for useful conversations.

\end{document}